\documentclass[3p,preprint]{elsarticle}

\usepackage{lineno,hyperref}
\modulolinenumbers[5]
\usepackage{amsmath,amssymb,amsthm}
\usepackage{graphicx}
\usepackage{epstopdf}
\usepackage{subfigure}
\DeclareMathOperator{\Tr}{Tr}
\DeclareMathOperator{\e}{e}

\makeatletter
\def\ps@pprintTitle{%
 \let\@oddhead\@empty
 \let\@evenhead\@empty
 \def\@oddfoot{\centerline{\thepage}}%
 \let\@evenfoot\@oddfoot}
\makeatother

\begin{document}

\begin{frontmatter}

\title{Construction of spin models displaying quantum criticality \\ from quantum field theory}

\author[mpq,tum]{Ivan Glasser}
\author[mpq]{J. Ignacio Cirac}
\author[madrid]{Germ\'an Sierra}
\author[mpq]{Anne E. B. Nielsen}
\address[mpq]{Max-Planck-Institut f{\"u}r Quantenoptik,
Hans-Kopfermann-Strasse 1, D-85748 Garching, Germany}
\address[madrid]{Instituto de F\'isica Te\'orica, UAM-CSIC, Madrid, Spain}
\address[tum]{Fakult{\"a}t f{\"u}r Physik, Technische Universit{\"a}t M{\"u}nchen, James-Franck-Str. 1, D-85748 Garching, Germany}


\begin{abstract} 
We provide a method for constructing finite temperature states of one-dimensional spin chains displaying quantum criticality. These models are constructed using correlators of products of quantum fields and have an analytical purification. Their properties can be investigated by Monte-Carlo simulations, 
which enable us to study the low-temperature phase diagram and to show that it displays a region of quantum criticality. The mixed states obtained are shown to be close to the thermal state of a simple nearest neighbour Hamiltonian.
\end{abstract}

\end{frontmatter}

\section{Introduction}

Not driven by thermal but by quantum fluctuations, quantum phase transitions occur in systems at zero temperature. Although matter can never be cooled down to the absolute zero temperature, the presence of a quantum critical point at zero temperature can strongly affect the low temperature phase diagram of a material \cite{sachdev,coleman,Sachdev2011}. Such behaviour has been observed in several experiments \cite{Coldea2010,Rueegg2008,Custers2003,Loehneysen1994,Grigera2001,Julian1996}, in which a solid is brought close to a critical point by tuning, for instance, an externally applied magnetic field, the pressure or the chemical composition of the material. Quantum critical points in one dimension are quite well understood due to their description in terms of conformal field theories (CFT)\cite{cft}, but it is harder to understand the complete region affected by the critical point : one has to take into account the strongly-coupled dynamics of the quantum critical point and its non-trivial excitations. There are, however, a few exactly solvable models that can be studied in great detail \cite{sachdev}.

It has been demonstrated \cite{cftimps,Nielsen2011,Tu2014,Tu2013} that a number of quantum spin states at zero temperature constructed from correlators of products of conformal fields display critical behaviour in one dimension. In the present paper, we propose to use this construction as a starting point for building models with a quantum critical region in their low temperature phase diagram. The first step in the construction is to introduce a parameter in the states that allows us to drive the states away from the critical point at zero temperature. This is done in a natural way by using massive fields rather than massless conformal fields to build the wave functions. The second step is to introduce a temperature, which is done by doubling the number of spins and considering the reduced state of half of the spins. There are very few models of quantum many-body systems for which an explicit parametrization of the density operator is known and can be used to determine physical properties numerically, but in this approach the analytical form of the considered state is known at least for a number of models. Furthermore, it allows us to investigate the properties of the quantum critical region with Monte Carlo techniques, so that quite large systems can be considered.

This prescription can be applied to construct models with different CFTs as a starting point. To illustrate our ideas, we shall in this work carry out the construction explicitly for the case where the CFT is a free massless boson and the corresponding initial CFT state is the wave function introduced in \cite{cftimps}. This state describes a critical spin chain close to the ground state of an anisotropic XXZ chain. We provide the analytical expression for the state with massive fields and study how the presence of the quantum critical point at zero temperature has an influence on the mixed state at non-zero temperature by computing two-point correlation functions and the Renyi mutual information.

It is also relevant to ask whether the states obtained from the above construction are described by a realistic Hamiltonian. In more cases, it has turned out that the states obtained from conformal fields have large overlaps with the ground states of few-body local Hamiltonians \cite{cftimps,Nielsen2013}, and one could therefore hope that this is also the case after introducing a mass and a non-zero temperature. For the example that we investigate in detail in the present work, we find that it is indeed possible to find a simple nearest neighbour Hamiltonian whose thermal state is close to our mixed state ansatz. This facilitates an experimental realization of this model.

The paper is organized as follows: In section II, the infinite dimensional MPS construction used for this work is introduced. In section III this ansatz is applied to a spin chain at zero temperature and its properties are characterized. In section IV the non-zero temperature case is investigated and the phase diagram is drawn. In section V it is shown that the investigated mixed state has a high fidelity with the thermal state of a nearest neighbour two-body Hamiltonian. Section VI concludes the paper. 

\section{Infinite dimensional MPS construction}

Let us consider $N$ spin-$1/2$ on a 2D lattice with local spin basis $|s_i \rangle$, $s_i= \pm 1$. In this work, $N$ will always be even. In this section we start by reviewing the infinite dimensional MPS construction that leads to a critical CFT wave function \cite{cftimps} from vertex operators of a free massless boson. We then introduce a parameter that takes the state away from criticality by generalizing this construction to the case of a free boson with mass $m$, thus obtaining a new wave function that reduces to the CFT wave function when $m=0$. \\

In general, a state in the Hilbert space can be written as
\begin{eqnarray}
|\psi \rangle = \sum_{s_1, \dots, s_N} \psi(s_1, \dots, s_N) \; |s_1, \dots, s_N \rangle,
\label{psi}
\end{eqnarray}
where $\psi(s_1, \dots, s_N)$ are complex coefficients. Infinite dimensional MPS are states for which these coefficients have the form of a vacuum expectation value of a product of vertex operators :
\begin{eqnarray}
\psi(s_1, \dots, s_N) \propto \langle \mathcal{V}_{s_1}(z_1) \dots \mathcal{V}_{s_N}(z_N) \rangle,\label{imps}
\end{eqnarray}
where the $z_n$ are complex numbers that represent the positions of the spins. \\

Let us first review the construction of a CFT wave function in the massless case. The vertex operators considered here are normal ordered exponentials of a field expressed as
\begin{eqnarray}
\mathcal{V}_{s_j}(z_j,\bar z_j)\equiv  : e^{i \sqrt{\alpha} s_j \phi(z_j,\bar z_j)} :,
\end{eqnarray}
where $\alpha$ is a real positive number and $\phi$ is the field of a free boson defined on a cylinder of circumference $R$:
\begin{align}
\phi(z_j,\bar z_j)=\varphi_0-i\pi_0 \ln(z \bar z)+\sum_{n=1}^\infty\frac{i}{\sqrt{n}} \left(a_n z_j^{-n} - a_n^\dag z_j^{n}\right)+\sum_{n=1}^\infty\frac{i}{\sqrt{n}} \left(\bar a_n \bar z_j^{-n} - \bar a_n^\dag \bar z_j^{n}\right),
\end{align}
where the non-zero commutators are $\left[a_n,a_m^\dag\right]=\left[\bar a_n,\bar a_m^\dag\right]=\delta_{nm}$, $\left[\varphi_0,\pi_0\right]=i$. The coordinates can alternatively be expressed as $z_n=e^{-2\pi i(x_n-it_n)/R}$ and $\bar z_n=e^{2\pi i(x_n+it_n)/R}$ in Euclidean space. Using the mode expansion of the field, the full correlator can be written as
\begin{align}
\left<:\e^{i \sqrt{\alpha} s_1 \phi(z_1,\bar z_1)} :\ldots :\e^{i \sqrt{\alpha} s_N \phi(z_N,\bar z_N)}:  \right> = \delta_s \prod_{j<k} \left| z_j-z_k \right|^{2\alpha s_j s_k},
\end{align}
where $\delta_s=1$ if $ \sum_{i=1}^N s_i =0$ and zero otherwise. From this expression we can define a wave function $\psi_0$  by imposing that the diagonal elements of the density operator only depend on the full correlator :
\begin{align}
\left| \psi_0 (s_1,\ldots,s_n) \right|^2 \propto \left< :\e^{i \sqrt{\alpha} s_1 \phi(z_1,\bar z_1)} :\ldots :\e^{i \sqrt{\alpha} s_N \phi(z_N,\bar z_N)}:  \right> .\label{fullcorrel}
\end{align}
This can be done by dividing the full correlator into
\begin{align} 
\psi_0 (s_1,\ldots,s_n) &\propto \prod_{i=1}^{N} \chi_i \ \delta_s \prod_{j<k} (z_j-z_k)^{\alpha s_j s_k}, \label{wavezeromass}
\end{align}
where the $\chi_i$ are phase factors to specify. This wave function is the CFT wave function introduced in \cite{cftimps}. Note that $\psi_0$ can be obtained by dividing the field $\phi$ into chiral fields $\varphi_L, \varphi_R$ such that $\phi(z,\bar{z})=\varphi_L(\bar{z})+\varphi_R(z)$  and then taking the correlator of vertex operators of the chiral field :
\begin{align}
\psi_0 (s_1,\ldots,s_n) &\propto \langle \chi_1 : e^{i \sqrt{\alpha} s_1 \varphi_R(z_1)} : \dots \chi_N : e^{i \sqrt{\alpha} s_N \varphi_R(z_N)} : \rangle.
\end{align}
The choice of phases that will be assumed here is
\begin{align}
\chi_n=(-1)^{(n-1)(s_n+1)/2}.\label{chiphases}
\end{align}
This choice ensures that the wave function is a singlet at $\alpha=1/2$ \cite{Nielsen2012}. The CFT wave function is expected to describe a critical state in 1D\cite{cftimps}. \\

We would like to introduce a parameter that takes the state away from criticality so we now apply the previous derivation to the case of a free boson $\phi_m$ with mass $m$, that breaks the conformal invariance. The mode expansion of the field $\phi_m$ is
\begin{align}
\phi_m(z_j,\bar z_j)=& i \sqrt{\frac{2\pi}{R m}} \left(a_0 z_j^{-\frac{m R}{4 \pi}} \bar z_j^{-\frac{m R}{4 \pi}} - a_0^\dag z_j^{\frac{m R}{4 \pi}} \bar z_j^{\frac{m R}{4 \pi}}\right) \nonumber \\
&+ i \sum_{n=1}^\infty\sqrt{\frac{2\pi}{Rw_n}} \left(a_n z_j^{-\frac{n}{2}-\frac{w_n R}{4 \pi}} \bar z_j^{\frac{n}{2}-\frac{w_n R}{4 \pi}} - a_n^\dag z_j^{\frac{n}{2}+\frac{w_n R}{4 \pi}} \bar z_j^{-\frac{n}{2}+\frac{w_n R}{4 \pi}}\right)\nonumber \\
&+ i \sum_{n=1}^\infty\sqrt{\frac{2\pi}{Rw_n}} \left(\bar a_n z_j^{+\frac{n}{2}-\frac{w_n R}{4 \pi}} \bar z_j^{-\frac{n}{2}-\frac{w_n R}{4 \pi}} - \bar a_n^\dag z_j^{-\frac{n}{2}+\frac{w_n R}{4 \pi}} \bar z_j^{\frac{n}{2}+\frac{w_n R}{4 \pi}}\right),
\end{align}
where the frequency is $w_n=\sqrt{m^2+\left(\frac{2 \pi n}{R}\right)^2}$.
The full correlator then becomes
\begin{align}
\left< :\e^{i \sqrt{\alpha} s_1 \phi_m(z_1,\bar z_1)} :\ldots :\e^{i \sqrt{\alpha} s_N \phi_m(z_N,\bar z_N)}:  \right>& = \nonumber \\
 \exp\left(-\alpha\sum_{j<k}s_j s_k \frac{2\pi}{Rm}\left(\frac{|z_k|}{|z_j|}\right)^{\frac{Rm}{2\pi}}      \right)
\exp & \left(-\alpha\sum_{j<k}s_j s_k \sum_{n=1}^{\infty} \frac{2\pi}{Rw_n} \left(\frac{z_k}{z_j}\right)^{\frac{n}{2}+\frac{w_n R}{4\pi}}   \left(\frac{\bar z_j}{\bar z_k}\right)^{\frac{n}{2}-\frac{w_n R}{4\pi}}     \right) \nonumber \\
\times \exp & \left(-\alpha\sum_{j<k}s_j s_k \sum_{n=1}^{\infty} \frac{2\pi}{Rw_n} \left(\frac{\bar z_k}{\bar z_j}\right)^{\frac{n}{2}+\frac{w_n R}{4\pi}}   \left(\frac{ z_j}{ z_k}\right)^{\frac{n}{2}-\frac{w_n R}{4\pi}}     \right). \label{fullcorrelatormass}
\end{align}
To obtain a wave function satisfying (\ref{fullcorrel}), one has to divide the correlator, which can be directly done by taking a wave function proportional to
\begin{align}
\exp\left(-\alpha\sum_{j<k}s_j s_k \frac{\pi}{Rm}\left(\frac{|z_k|}{|z_j|}\right)^{\frac{Rm}{2\pi}}      \right)
\exp & \left(-\alpha\sum_{j<k}s_j s_k \sum_{n=1}^{\infty} \frac{2\pi}{Rw_n} \left(\frac{z_k}{z_j}\right)^{\frac{n}{2}+\frac{w_n R}{4\pi}}   \left(\frac{\bar z_j}{\bar z_k}\right)^{\frac{n}{2}-\frac{w_n R}{4\pi}}     \right).
\end{align}
However this expression is not translational invariant when the spins are placed on a circle at positions $z_j=\e^{2\pi i j/N}$. To solve this problem, we take the limit $R\rightarrow \infty$ in the previous expression. This corresponds to a change in geometry in which the cylinder becomes a plane, so that a chain of spins on the circle with periodic boundary conditions becomes an open line of spins on the plane. This leads to the wave function
\begin{align}
\psi_m(s_1, \dots, s_N) & \propto \prod_{i=1}^{N} \chi_i \exp \left(-\alpha \sum_{j<k} s_j s_k \int_0^{\infty} \frac{ e^{- i p (x_k-x_j)}}{\sqrt{m^2+p^2}}e^{- \sqrt{m^2+p^2} (t_k-t_j)}\text{d}p\right).
\label{wave}
\end{align}
This wave function is such that the diagonal elements of the density operator are equal to the full correlator for the field with mass (\ref{fullcorrelatormass}). In the limit $m \rightarrow 0$, the wave function reduces to 
\begin{align}
\psi_0(s_1, \dots, s_N) & \propto \delta_s \prod_{j=1}^{N} \chi_j \prod_{j<k} (v_j-v_k)^{\alpha s_j s_k},
\label{wave0}
\end{align}
where $v_j=x_j-i t_j$. Thus in the massless case we recover, in the plane geometry, the CFT wave function (\ref{wavezeromass}) for the same particular choice of phases (\ref{chiphases}) that will be assumed in the rest of this work.

\section{Quantum critical point at zero temperature} 

Let us now apply this wave function to the case of a 1-dimensional chain of $N$ spins at positions $x_j=j$, $t_j=0$ with open boundary conditions. The integral in the wave function can be computed once expressed as :
\begin{align}
\int_0^{\infty} \frac{ e^{-i p (x_k-x_j)}}{\sqrt{m^2+p^2}}\text{d}p = K_0 (m|x_k-x_j|)-\frac{i \pi}{2} \frac{k-j}{|k-j|} \left(I_0 (m|x_k-x_j|) -L_0(m|x_k-x_j|)\right),
\end{align}
where $K_0$ and $I_0$ are modified Bessel functions and $L_0$ is a modified Struve function. The wave function depends only on the distance between two spins but is not translational invariant because of the choice of phase factors. In the thermodynamic limit, the wave function becomes invariant under a translation $x_k\rightarrow x_k+2$. We shall only consider a value of $\alpha$ in the range $(0,1/2]$. In this case when the mass is zero and the spins are on a chain with periodic boundary conditions, the wave function corresponds to the CFT wave function (\ref{wavezeromass}) studied in \cite{cftimps} : the state is critical and close to the ground state of a critical XXZ chain. Here we study a chain with open boundary conditions but we expect the state to have similar properties in the massless limit.

\begin{figure}[h]
\begin{center}
  \subfigure[]{\includegraphics[scale=0.61]{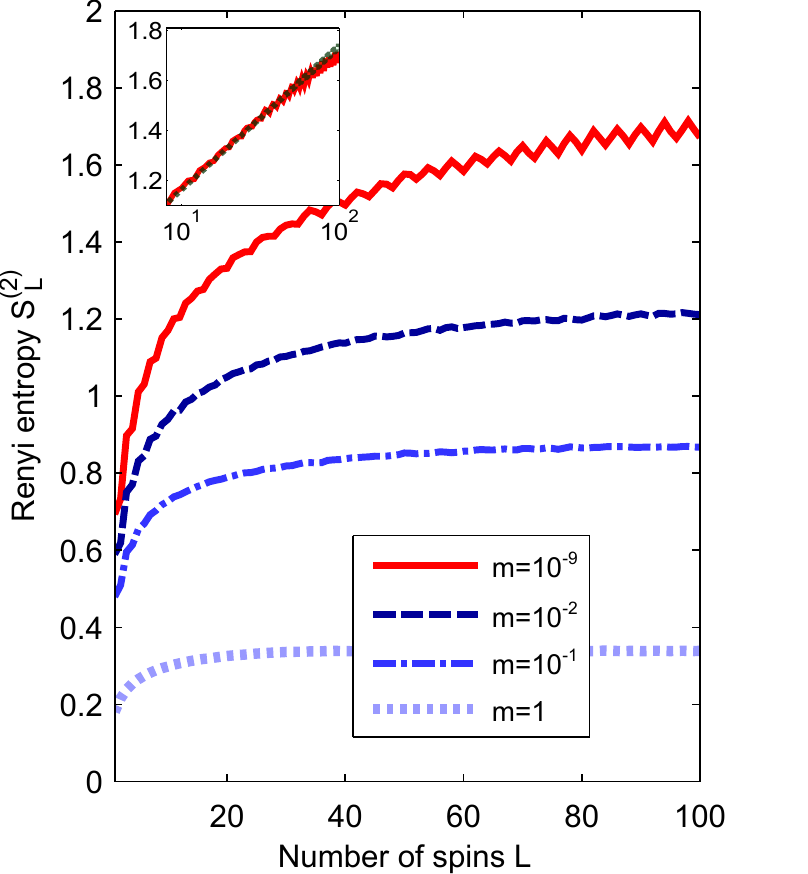}\label{sub1a}}
  \subfigure[]{\includegraphics[scale=0.62]{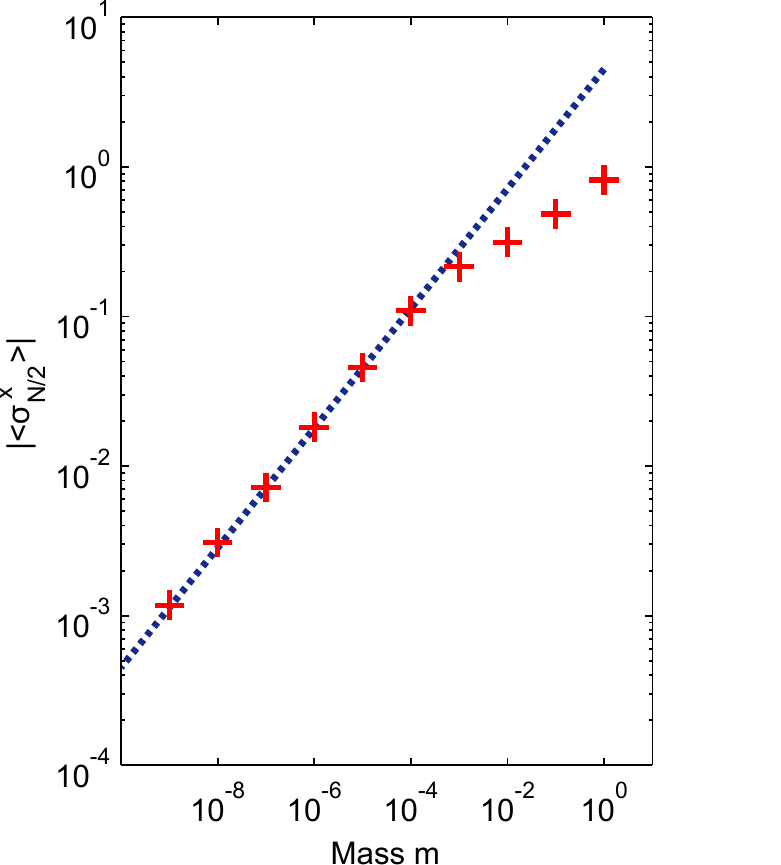}\label{sub1b}}
  \subfigure[]{\includegraphics[scale=0.61]{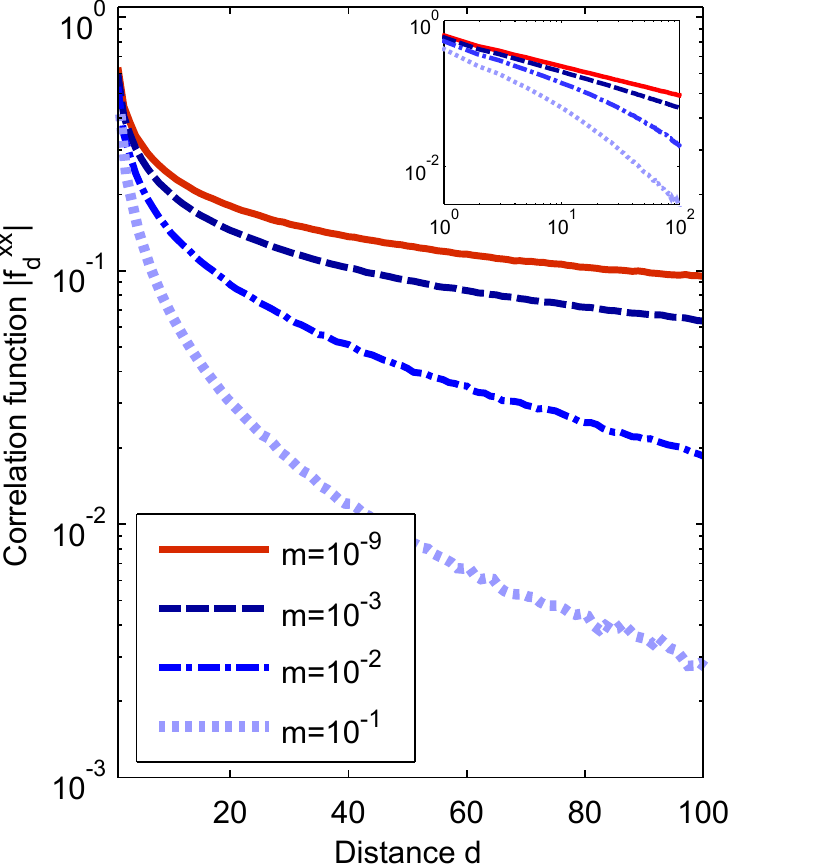} \label{sub1c}}
\caption{All quantities are evaluated from Monte-Carlo simulations for $\alpha=0.2$. \newline
(a) Scaling of the Renyi entropy $S^{(2)}_L$ for different values of the mass, $N=200$ spins. The inset shows with a logarithmic scale for the abscissa that the scaling is logarithmic in $L$ when the mass is close to zero. The dotted line in the inset is a fit of the form $\frac{1}{4}\log(L)+\text{constant}$, confirming that the central charge of the critical point is 1. For higher masses, the entropy saturates to a constant. The error bars, estimated from the standard error of a mean of Monte-Carlo trajectories with different initial conditions, are of the order $10^{-3}$ for all points.\newline
(b) Absolute value of the magnetization in the $x$ direction for one site $\langle \sigma^x_{N/2} \rangle$ as a function of the mass for a chain with $N=600$ spins. The error bars are smaller than $5\%$ of the values for all points. The dotted line is a fit of the first 6 points of the form $\gamma m^\tau$, from which the critical exponent can be extracted. Here $\tau=0.4$.\newline
(c) Absolute value of the connected correlation function $f^{xx}_d$, $N=200$ spins. When the mass is zero this quantity decays polynomially (red solid line), while in the massive case the decay for large d is exponential. The inset shows the same quantity in a log-log scale. The error bars are smaller than $2\times 10^{-4}$ for all points.}
\end{center}
\end{figure}

We compute the Renyi entropy $S^{(2)}_L = - \log {\rm Tr} \;  \rho^2_L$, where $\rho_L$ is the density matrix of (\ref{wave}) restricted to a subsystem of size $L$ in the middle of the chain. This can be done \cite{cftimps,Hastings2010} by rewriting
 \begin{equation}
 e^{-S^{(2)}_L} = \sum_{n,n',m,m'}
 |\langle n,m|\psi\rangle|^2 |\langle n',m'|\psi\rangle|^2
 \frac{ \langle\psi|n',m\rangle \langle\psi|n,m'\rangle}
 {\langle\psi|n,m\rangle \langle\psi|n',m'\rangle} \left[\sum_{n,m} |\langle n,m|\psi\rangle|^2\right]^{-2},
 \end{equation}
where $|n\rangle$ (and $|n'\rangle$) is an orthonormal basis in the space of the $L$ spins and $|m\rangle$ (and $|m'\rangle$) another basis corresponding to the rest of spins. The right-hand side of this expression is an expectation value so that the sum can be performed by using a Metropolis-Hastings algorithm \cite{Metropolis1953,Hastings1970} with two independent spin chains. The results for different values of the mass are shown in Fig. \ref{sub1a}. When the mass is close to zero, the second Renyi entropy scales as $\frac{1}{4}\log(L)$, which is the result expected for an infinite critical chain with central charge $c=1$\cite{c1,c2,c3}. For higher masses, the entropy saturates to a constant that is independent of $L$, so that the state is no longer critical.

Other quantities that can be computed using Monte-Carlo techniques are expectation values of single spin operators $\langle \sigma^a_n \rangle \; (a=x,y,z)$ and two-point correlation functions $f^{aa}_d = \langle \sigma^a_n \sigma^a_{n+d} \rangle - \langle \sigma^a_n \rangle \langle \sigma^a_{n+d} \rangle $. Since we are interested in the thermodynamic limit, we compute the correlators between spins at positions $(N-d)/2$ and $(N+d)/2$ that sit in the middle of the chain. We check that these quantities do not depend on the total number of spins $N$ as long as $d<N/2$, so that the behaviour in the thermodynamic limit can be extracted from these measurements. The results in Fig. \ref{sub1b} and \ref{sub1c} show that in the massless case $f^{xx}_d$ decays polynomially and the expectation value $\langle \sigma^x_n \rangle$ is zero. In the massive case however this expectation value is no longer zero, but shows long-range anti-ferromagnetic order in the x direction. The correlation function $f^{xx}_d$ decays exponentially at large distances when there is a mass : this defines a finite correlation length $\lambda$ such that $f^{xx}_d \propto\e^{-d/\lambda}$. This length diverges when the mass goes to zero, while in the limit $m \rightarrow \infty$, the state is a N\'{e}el state in the $x$ direction, which is invariant under translations $x_k\rightarrow x_k+2$. The mass therefore introduces a length scale in the system and breaks the criticality of the state.


\section{Phase diagram at non-zero temperature}

\begin{figure}[h]
\begin{center}
\includegraphics[scale=0.45]{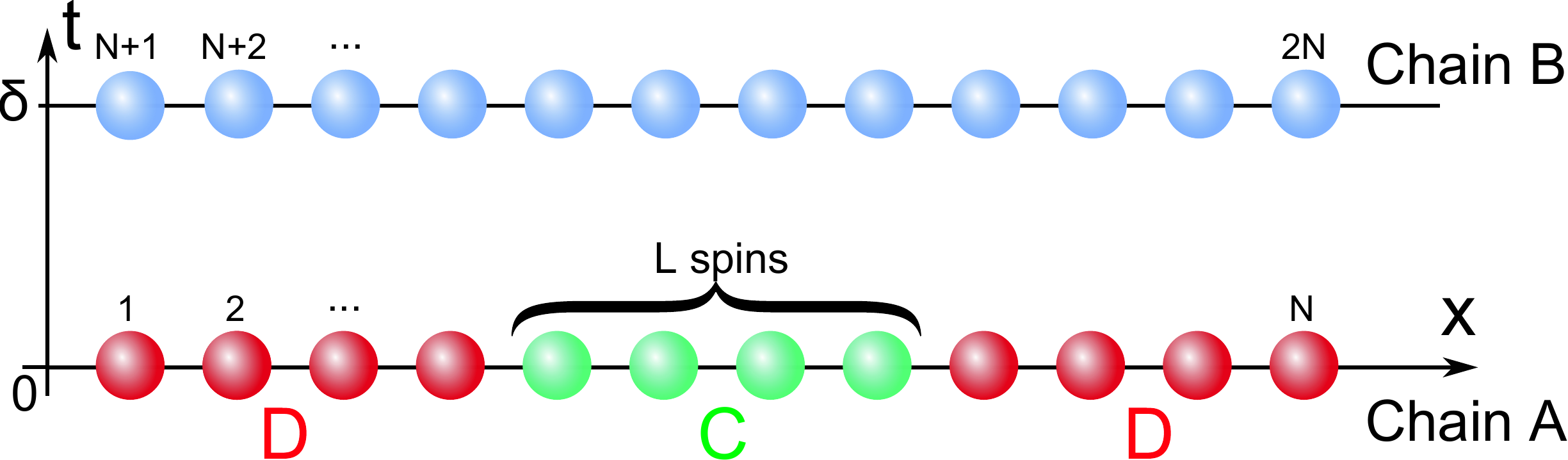}
\caption{The two spin chains A and B are separated by a distance $\delta$. The one dimensional state that is studied is the reduced state of the chain A. To compute the mutual information we separate the chain into the systems C of $L$ spins in the middle of the chain and the complementary system D of $N-L$ spins.}
\label{2chains}
\end{center}
\end{figure}

So far we have used a pure state description at zero temperature. Let us now introduce a mixed state ansatz to describe a spin chain at finite temperature. Consider two chains A and B of $N$ spins each, with coordinates ($x_j=j$, $t_j=0$) and ($x_j=j$, $t_j=\delta$) respectively (Fig. \ref{2chains}). Let us describe the state of the complete system by the previous wave function $\psi_{AB}$ (\ref{wave}). The state of the first spin chain is now given by the reduced density matrix $\rho_A=\Tr_B |\psi_{AB}\rangle\langle \psi_{AB}|$, where the trace is performed over the degrees of freedom of the second spin chain. In the limit where $\delta \rightarrow \infty$, the two chains decouple, so that the system A is in a pure state at zero temperature : we recover two copies of the state described in the previous section. In the limit $\delta \rightarrow 0$, each spin from the first chain is very close to a spin from the second chain and they form a singlet, so that the effective temperature for one chain goes to infinity. For a finite $\delta$, this construction therefore introduces an effective temperature for the chain A. We define $T\equiv 1/\delta$ as a representation of the temperature of the system A. Note that the effective temperature may depend differently on $\delta$.

In general it is not possible to compute the complete wave function and take the partial trace for a large system, but it is not necessary in this case : Renyi entropies and spin-spin correlators between spins on the first chain can be computed using the wave function of the two chains in the same way as in the zero temperature case. 

\begin{figure}[h]
\begin{center}
\includegraphics[scale=0.60]{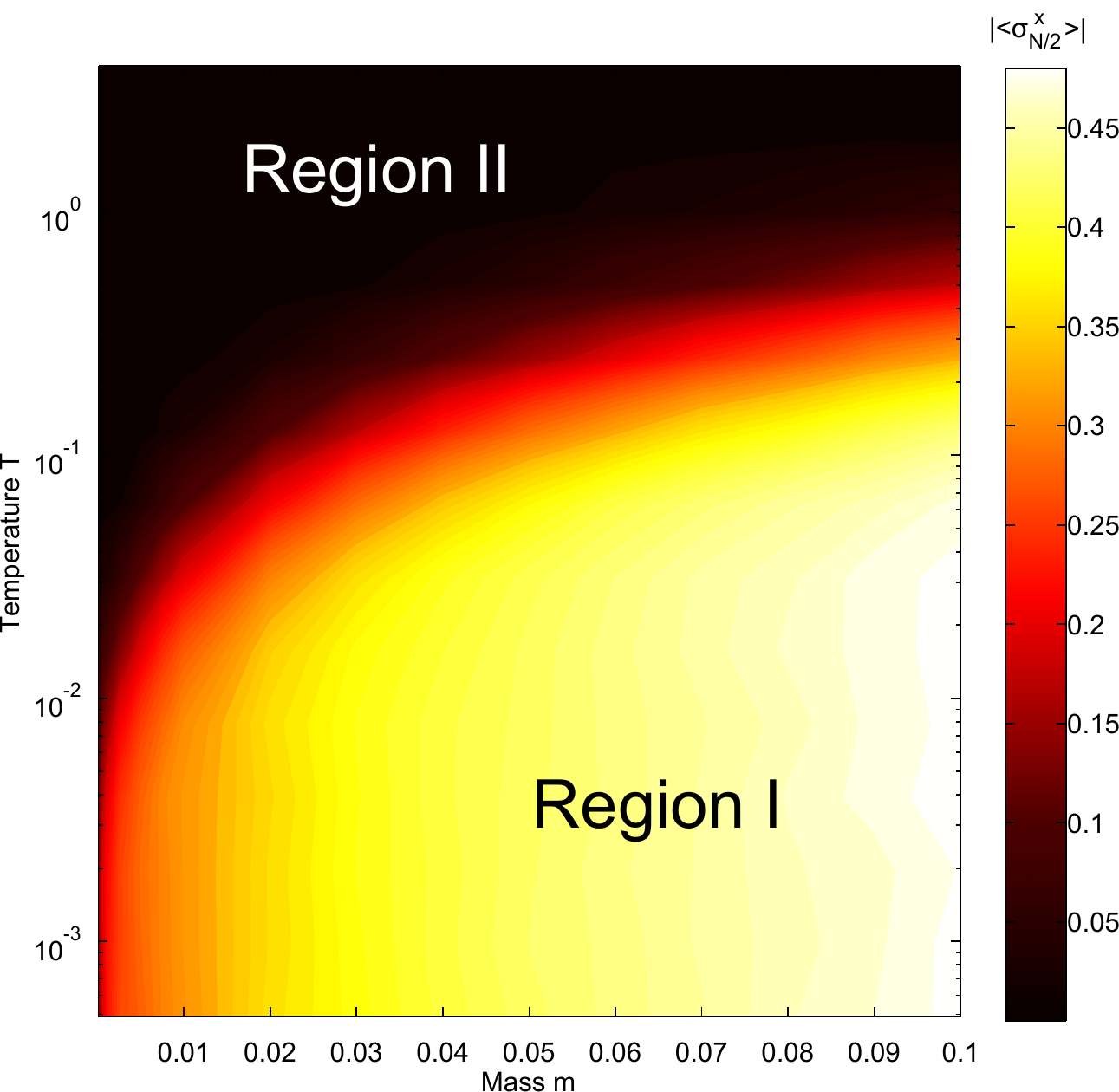}
\caption{Absolute value of the magnetization in the $x$ direction for one site $\langle \sigma^x_{N/2} \rangle$ as a function of mass and temperature from Monte-Carlo simulations with $N=600$, $\alpha=0.2$.}
\label{phase}
\end{center}
\end{figure}

\begin{figure}[h]
\begin{center}
\includegraphics[scale=0.65]{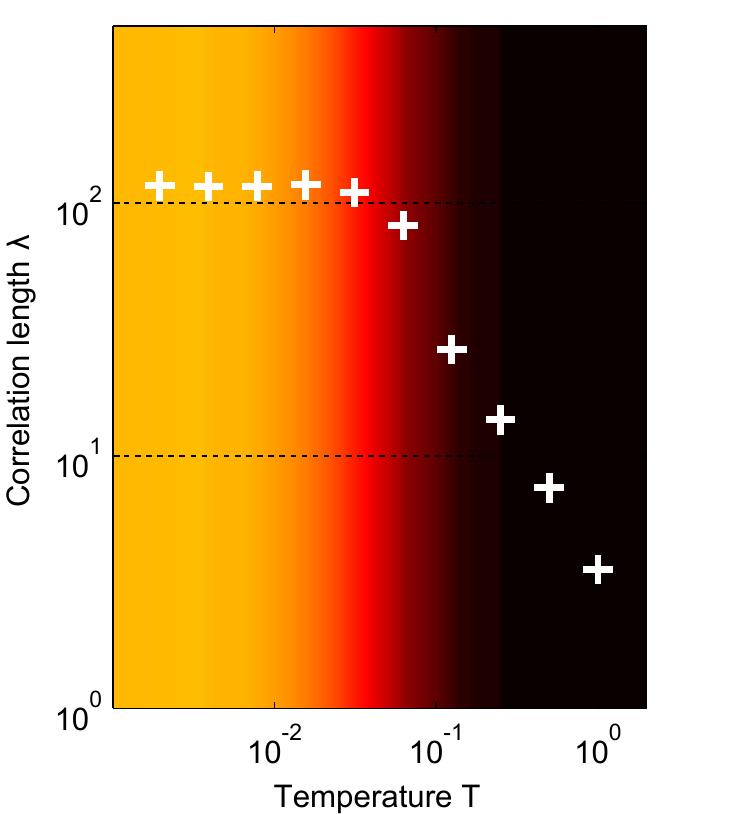}
\caption{Correlation length $\lambda$ as a function of the temperature for a spin chain with $N=200$ and $\alpha=0.2$, at $m=0.01$. The background colours represent the value of $|\langle \sigma^x_{N/2} \rangle|$ taken from the phase diagram (Fig. \ref{phase}) for the same value of the mass.}
\label{corrlength}
\end{center}
\end{figure}

At zero temperature and in the thermodynamic limit, the expectation value $\langle \sigma^x_i \rangle$ is zero at the critical point, but does not vanish when there is a mass. The absolute value of this quantity, computed on a site at position $i=\frac{N}{2}$ in the middle of a chain of 600 spins, is used to draw the phase diagram at finite temperature (Fig. \ref{phase}). Two distinct regions appear in the phase diagram : for small temperatures and non-zero mass, $|\langle \sigma^x_{N/2} \rangle|$ is non-zero and independent of the temperature, there is still long-range order in the $x$ direction (region I). At some higher temperature, the magnetization starts to decrease rapidly with the temperature, before reaching a very small value, which may disappear in the thermodynamic limit (region II). These two distinct regions also appear when looking at the correlation length $\lambda$ (Fig. \ref{corrlength}) : for small temperatures, the correlation length is independent of the temperature, whereas in the second region the correlation length decreases with the temperature. Such a behaviour can be qualitatively compared with the phase diagram of an Ising model with a transverse magnetic field, which is the prototype model of quantum criticality \cite{Pfeuty1970,Elliott1970,sachdev} : this model has a quantum paramagnetic phase at low temperatures in which the correlation length is independent of the temperature, while at higher temperatures it reaches a region of quantum criticality in which the correlation length decays as $1/T$.


%

\begin{figure}[h]
\begin{center}
  \subfigure[]{\includegraphics[scale=0.58]{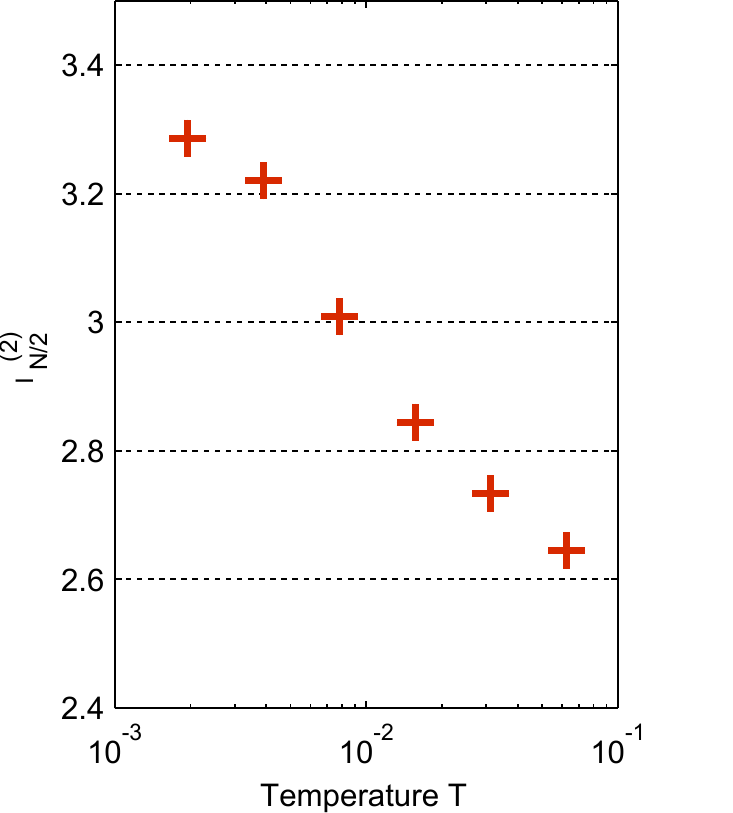}\label{entropytempzero}}
  \subfigure[]{\includegraphics[scale=0.55]{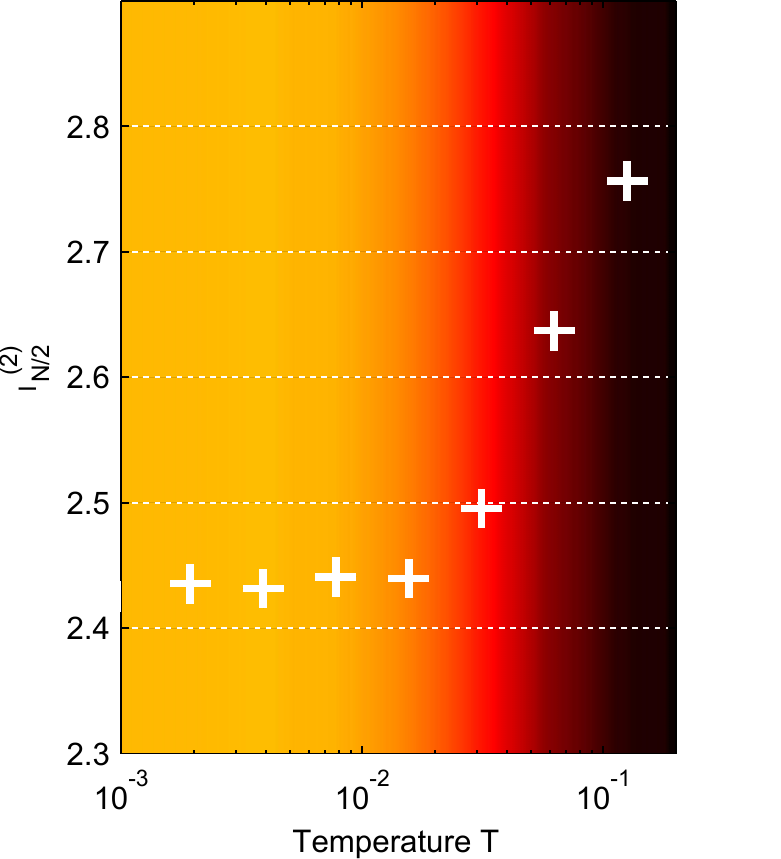}\label{entropytemp}}
\caption{(a) Mutual information between two halves of the chain $I_{N/2}^{(2)}$ for $N=200$, $\alpha=0.2$ and $m=10^{-6}$ : for this value of the mass we are almost above the quantum critical point and the mutual information decreases as we go away from the critical point by increasing the temperature. The error bars, estimated from the standard error of a mean of Monte-Carlo trajectories with different initial conditions, are smaller than $2\times 10^{-2}$ for all points. \newline
(b) Mutual information between two halves of the chain $I_{N/2}^{(2)}$ for $N=200$, $\alpha=0.2$ and $m=0.01$. The background colours represent the value of $|\langle \sigma^x_{N/2} \rangle|$ taken from the phase diagram (Fig. \ref{phase}) for the same value of the mass : the mutual information increases as we enter the region of quantum criticality. The error bars are smaller than $5\times 10^{-3}$ for all points.}
\end{center}
\end{figure}

Another quantity that can be used to probe a state at finite temperature is the mutual information $I=S_C+S_{D}-S_{C \cup D}$, where $\{C,D\}$ is a partition of the spin chain and $S_C$ is the von Neumann entropy of subsystem C. The mutual information is a measure of both classical and quantum correlations between two parts of the system \cite{Groisman2005} and it fulfils an area law when the system has a finite correlation length \cite{Wolf2008}. Here we consider instead the Renyi mutual information, obtained by replacing von Neumann entropies by Renyi entropies in the definition of the mutual information : $I^{(2)}_L=S^{(2)}_C+S^{(2)}_{D}-S^{(2)}_{A}$, where C is the system of $L$ spins in the middle of the chain and D is the system of the $N-L$ complementary spins in the chain A (Fig. \ref{2chains}). This quantity has been measured for spin systems at finite temperature \cite{Melko2010,Singh2011,Bonnes2013} and is expected to have a similar behaviour as the mutual information. When the chain is in a pure state (at zero temperature) the entropy of the complete chain is zero so the Renyi mutual information reduces to $2S^{(2)}_L$. At the critical point this quantity therefore scales logarithmically with $L$. Except when we are close to the critical point, we observe that the Renyi mutual information saturates to a constant at large $L$, which is compatible with the observation that the correlation length is finite. In region I this constant is independent of the temperature. In region II however the Renyi mutual information converges to a value that has some non trivial dependence on the temperature. Above the critical point the Renyi mutual information between two halves of the system decays with the temperature (Fig. \ref{entropytempzero}), which is expected for an XXZ chain at finite temperature \cite{Znidaric2008} since the mutual information diverges at zero temperature. In the massive case this quantity increases when the region of quantum criticality is reached (Fig. \ref{entropytemp}), while in the limit of infinite temperatures there would be no correlations and the mutual information would decay to zero. A similar behaviour of the mutual information has been observed near regions of quantum criticality at finite temperature in different models \cite{Wilms2012,Melko2010,Bonnes2013}. These results confirm the presence of a region of quantum criticality above the critical point in the phase diagram.



\section{Hamiltonian of the model}

The system considered here may correspond to a non-local Hamiltonian. However, the fact that the mutual information saturates at finite temperature may be a hint that this state could be a thermal state of a local Hamiltonian. For a general Hamiltonian, a thermal state is given by 
\begin{align}\rho_{th} = e^{-\beta H}/\Tr \left[e^{-\beta H}\right]\label{rho}.\end{align}
To investigate the Hamiltonian of the system we therefore define
\begin{eqnarray}
H_{\rho_A} = - \log \rho_A, \label{Hrho}
\end{eqnarray}
where $\rho_A$ is the thermal state describing the chain A at a finite temperature. Note that the Hamiltonian may have a non-trivial dependence on the temperature and be non local. However it was shown in \cite{cftimps} for periodic boundary conditions that the ansatz in the massless case with $\alpha\in\left[0,\frac{1}{2}\right]$ has a very high overlap with the exact ground state of the XXZ chain for a suitable choice of the anisotropic coupling. For open boundary conditions, the Hamiltonian of the XXZ chain is written as
\begin{eqnarray}
H_{XXZ} = \sum_{i=1}^{N-1} \left(  S^x_i S^x_{i+1} +   S^y_i S^y_{i+1}+ \Delta  S^z_i S^z_{i+1}\right),
\label{XXZ}
\end{eqnarray}
where $S^{\alpha}_i=\frac{1}{2}\sigma^{\alpha}_i$ for $\alpha \in \{x,y,z\}$ and the $\sigma^{\alpha}$ are Pauli matrices.

 In the case studied here of the wave function (\ref{wave}) for an open chain of spins, the correspondence with an XXZ chain still holds in the massless limit but breaks down in the presence of a mass. However we observe that when the mass is close to zero and the temperature is high, the Hamiltonian $H_{\rho_A}$ restricted to two-body interactions has the form of an $H_{XXZ}$ Hamiltonian, up to some non translational invariant corrections. This suggests to look at $H_{\rho_A}$ at high temperatures in the massive case. The Hamiltonian restricted to two-body interactions has, up to some non translational invariant terms, the form

\begin{align}
H_m&=H_{XXZ}+H^0_m \label{hamilto},\\
H^0_m&=\lambda \sum_{i=1}^N (-1)^i  S^x_i + \mu \sum_{j=1}^{N-1} S^z_j S^y_{j+1}.
\end{align}

In the thermodynamic limit, this Hamiltonian is invariant under translations $x_k\rightarrow x_k+2$, as is the wave function (\ref{wave}). Let us now define a thermal state $\rho_H$ for the Hamiltonian $H_m$ by :
\begin{align}\rho_H = e^{-\beta H_m}/\Tr \left[e^{-\beta H_m}\right]\label{rhoH}.\end{align}
This state depends on the parameters $\beta$, $\Delta$, $\lambda$ and $\mu$. A way to check whether this Hamiltonian can correspond to our system is to compute the fidelity \cite{fidelity,Jozsa1994} between this state and the state $\rho_A$ :
\begin{align}
F(\rho_H,\rho_A)=\Tr \left[\sqrt{\sqrt{\rho_H}\rho_A\sqrt{\rho_H}}\right].
\end{align}

\begin{figure}[h]
\begin{center}
 \subfigure[]{\includegraphics[scale=0.72]{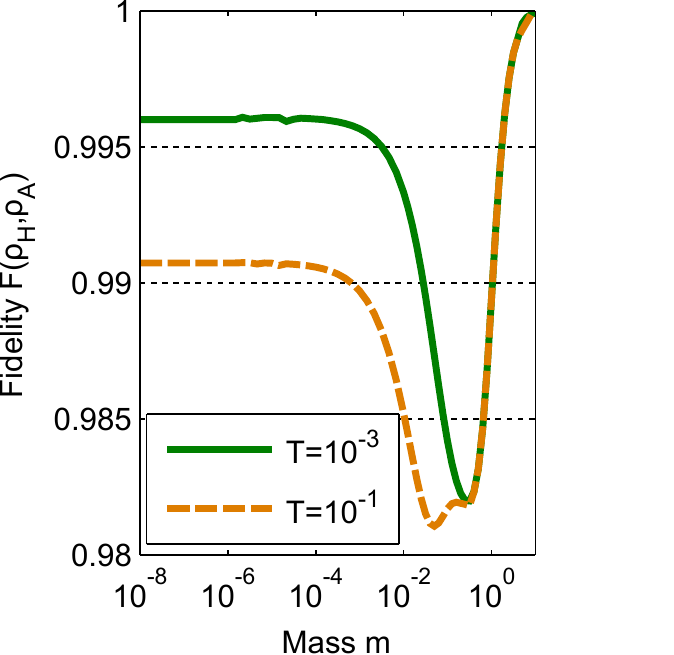}\label{subfidelitya}}
 \subfigure[]{\includegraphics[scale=0.61]{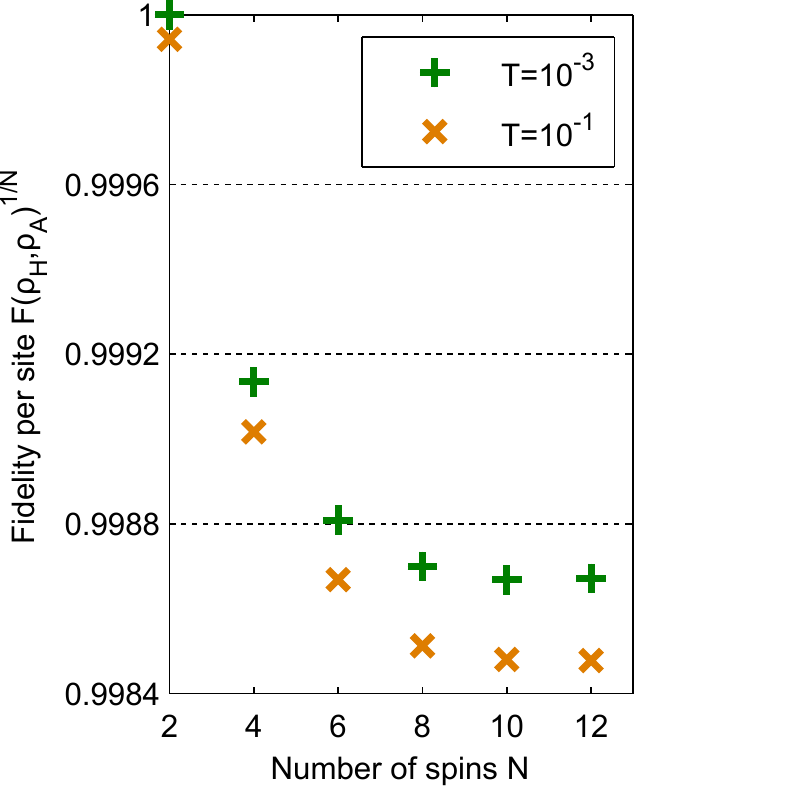}\label{subfidelityb}}
 \subfigure[]{\includegraphics[scale=0.62]{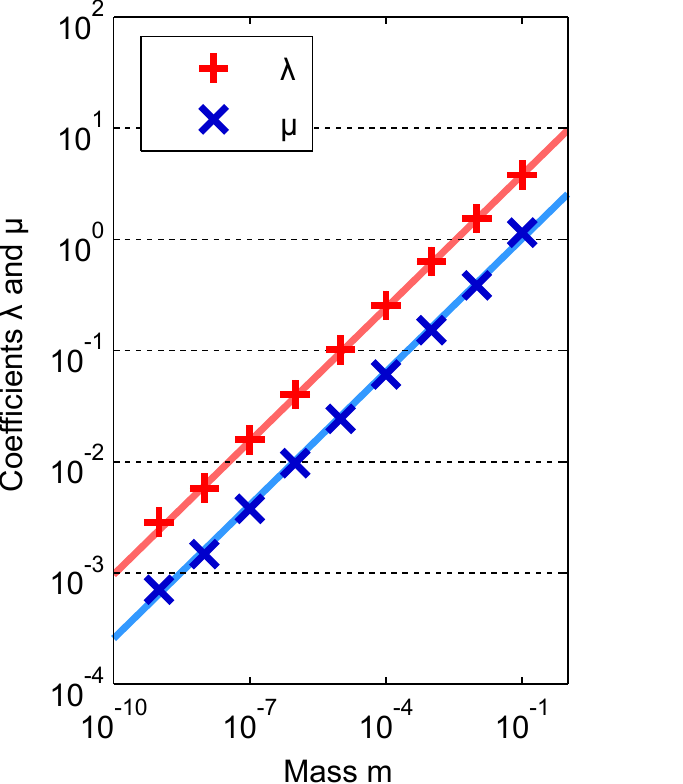}\label{scalingmu}}
\caption{(a) Maximum of the fidelity $F(\rho_H,\rho_A)$ between a thermal state obtained from a Hamiltonian of the form $H_m$ and the state of chain A as a function of the mass, at $\alpha=0.2$  and different temperatures ($N=12$ spins).\newline
(b) Corresponding fidelity per site $F(\rho_H,\rho_A)^{1/N}$ for different numbers of spins, at $\alpha=0.2$ and $m=10^{-1}$.\newline
(c) Parameters $\lambda$ and $\mu$ in the Hamiltonian $H_m$ for which the fidelity $F(\rho_H,\rho_A)$ is maximal as a function of the mass, for $\alpha=0.2$, $T=10^{-3}$, $N=10$. The two straight lines are fits of these data of the form $\lambda=\lambda_0m^{2\alpha}$, $\mu=\mu_0m^{2\alpha}$.}
\end{center}
\end{figure}

For different values of $\alpha$, $m$ and $T$ we can thus find parameters $\beta$, $\Delta$, $\lambda$ and $\mu$ such that the fidelity is maximized. The maximum for $N=12$ spins (24 spins in total for the two chains) is shown in Fig. \ref{subfidelitya} for different values of the mass and the temperature. When the mass is zero and the temperature is small, we recover the result from \cite{cftimps} : the fidelity is above $99\%$ when $\lambda$ and $\mu$ are zero, so that the state is close to the ground state of the XXZ chain. In the massive case the thermal state from Hamiltonian $H_m$ has a fidelity higher than $98\%$ with the state of chain A for all values of the mass and $T$ smaller than $0.1$, which is very high considering the size of the Hilbert space : the fidelity per lattice site $F^{1/N}$ is higher than $99.85\%$ (Fig. \ref{subfidelityb}). 

Note that further constraints can be imposed on the Hamiltonian by using the result in the massless case to fix the parameters $\beta$ and $\Delta$, that are related to $T$ and $\alpha$. By increasing the mass we then find that there exist two functions $\lambda(m)$ and $\mu(m)$ for which the previous result still holds. In the regime $m \in (0,10^{-1}],\ \alpha<1/4$, we note that these functions can be written as $\lambda(m)=\lambda_0 m^{2\alpha}$ and $\mu(m)=\mu_0 m^{2\alpha}$, where $\lambda_0$ and $\mu_0$ are constants independent of the mass (Fig. \ref{scalingmu}).
The ansatz (\ref{wave}) therefore can be used in this regime to describe the thermal state of a spin chain governed by a Hamiltonian of the form (\ref{hamilto}).
The two-body Hamiltonians found in this section contain only nearest neighbour interactions and should be easier to implement than the Hamiltonian $H_{\rho_A}$.

\section{Conclusion}
Using correlators of products of quantum fields, we have constructed a thermal state for a one-dimensional spin chain at finite temperature. This model has an analytical wave function and can be investigated by Monte-Carlo simulations, which enable us to study the entanglement properties and correlators of the spin chain. These quantities show that it presents a critical point as well as a non-critical phase at zero temperature. The phase diagram of this model has then been investigated and it was shown that it presents a region above the critical point which is the region of quantum criticality. We also provided a Hamiltonian with only nearest neighbour interactions whose thermal state is very close to this model, so that an experimental realisation of this model is facilitated.

\section*{Acknowledgements}

The authors acknowledge discussions with Maitagorri Schade. This work has been supported by the EU project SIQS, FIS2012-33642, QUITEMAD (CAM) and the Severo Ochoa Program.

\section*{References}

\biboptions{sort&compress}
\bibliography{mybibfile2}

\end{document}